\documentclass{PoS}
\usepackage{amsmath}
\usepackage{amssymb}

\title{Perturbative O($a^2$) effects in gradient flow couplings with SF and SF-open boundary conditions}

\ShortTitle{Perturbative O($a^2$) effects in gradient flow couplings}

\author{\speaker{Argia Rubeo} \\
        School of Mathematics and Hamilton Mathematics Institute, Trinity College Dublin, Dublin 2, Ireland\\
        E-mail: \email{rubeoar@maths.tcd.ie}}

\author{Stefan Sint\\
        School of Mathematics and Hamilton Mathematics Institute, Trinity College Dublin, Dublin 2, Ireland\\
        E-mail: \email{sint@maths.tcd.ie}}

\abstract{The gradient flow provides a new class of renormalized observables which 
can be measured with high precision in lattice simulations. In principle 
this allows for many interesting applications to renormalization and improvement problems.
In practice, however, such applications are made difficult by the rather large cutoff effects
found in many gradient flow observables. At lowest order of perturbation theory we here study the leading cutoff effects
in a finite volume gradient flow coupling with SF and SF-open boundary conditions. 
We confirm that O($a^2$) Symanzik improvement is achieved at tree-level,
provided the action, observable and the flow are O($a^2$) improved.
O($a^2$) effects from the time boundaries are found to be absent at this order, 
both with SF and SF-open boundary conditions.
For the calculation we have used a convenient representation of the free gauge field propagator at finite flow times which 
follows from a recently proposed set-up by L\"uscher and renders 
lattice perturbation theory more practical at finite 
flow time and with SF, open, SF-open or open-SF boundary conditions.}

\FullConference{34th annual International Symposium on Lattice Field Theory\\
		24-30 July 2016\\
		University of Southampton, UK}

\begin{document}

\section{Introduction}

Following the seminal work by L\"uscher~\cite{Luscher:2010iy} the Yang-Mills gradient flow has become a useful tool in lattice gauge
theories.  A whole new class of gauge invariant observables with simple renormalization properties~\cite{Luscher:2011bx}  has become available.
Moreover, high statistical precision can be achieved due to the smoothing properties of the gradient flow.
In principle this offers new approaches for difficult non-perturbative renormalization problems and to 
non-perturbative Symanzik improvement for lattice actions and fields~\cite{Luscher:2013vga}.
In practice, however, a major drawback has been the observation of rather large $a^2$-effects in many 
gradient flow observables, cf.~\cite{Ramos:2015dla}.
This motivated the application of the Symanzik programme to O($a^2$) in \cite{Ramos:2015baa} which
revealed that complete O($a^2$) improvement requires four ingredients: 
\begin{enumerate}
\item  O($a^2$) improvement of the lattice gauge action;
\item  an $a^2$- modification of the initial condition for the gradient flow equation at flow time $t=0$;
\item  classical O($a^2$) improvement of the flow equation and
\item  of the observable under study.
\end{enumerate}
For the last two ingredients Symanzik O($a^2$) improvement
can easily be implemented non-per\-tur\-bative\-ly~\cite{Ramos:2015baa}.  
Regarding the first two items on this list, we here use
the tree-level O($a^2$) improved L\"uscher-Weisz (LW) action 
and the initial condition then requires no modification to this order.
We here test this expectation using the finite volume coupling with SF
and open-SF boundary conditions at $x_0=0,T$ as a test case. Such boundary conditions
are particularly convenient for QCD, as they allow for simulations in the chiral
limit and for gauge invariant correlation functions with external fermion lines.
In particular, running gradient flow couplings with SF boundary conditions~\cite{Fritzsch:2013je} can thus be studied
directly in massless QCD.  Following~\cite{DallaBrida:2016kgh}  
we focus on the magnetic energy density and for SF boundary conditions 
our calculation confirms results obtained there.
Regarding the lattice action near the time boundaries at $x_0=0,T$, we use a
different set-up which has been proposed by L\"uscher in \cite{Luscher:2014kea}. As a result
a convenient representation of the free gauge field propagator is obtained,
and changing from SF to open boundary conditions or a mixture of these (SF-open or open-SF)
becomes rather trivial.


\section{The gradient flow in the continuum and on the lattice}

In the continuum, the Yang-Mills gradient flow equation
\begin{equation} 
\label{flowct}
\frac{\partial B_{\mu}  (t,x) }{\partial t}  = D_{\nu} G_{\nu \mu} (t,x),   \quad B_{\mu} (t,x)  |_{t=0} = A_{\mu} (x),
\end{equation}
defines a mapping from the fundamental gauge field, $A_{\mu}(x)$, to another gauge field $B_{\mu}(x,t)$, parameterized by 
the  flow time parameter $t\ge 0$. Here the covariant derivative $D_\mu = \partial_\mu + [B_\mu,\cdot]$ of the
field strength tensor,
\begin{equation}
  G_{\mu \nu}(t,x)  = \partial_{\mu} B_{\nu}(t,x)  - \partial_{\nu} B_{\mu}(t,x) + [B_{\mu}(t,x) , B_{\nu}(t,x)] \,,
\end{equation}
arises as the gradient of the Yang-Mills action, so that the $B$-field is driven 
towards the smooth classical minimum of the Yang-Mills action.
A most remarkable property of the gradient flow is the fact that
gauge invariant composite fields at finite flow time, such as the colour 
magnetic energy density\footnote{We define e.g. $B_\mu = B_\mu^a T^a$ with anti-hermitian generators 
$T^a$ of SU($N$) normalized to ${\rm tr}(T^aT^b)=-\frac12 \delta^{ab}$. Colour indices take values $a=1,\ldots,N^2-1$, 
and a summation convention is assumed.},
\begin{equation}
  E_{\rm mag}(t,x) = -\frac12 \sum_{k,l=1}^3{\rm  tr}\{G_{kl}(t,x)G_{kl}(t,x)\} =  \frac14 \sum_{k,l=1}^3 G^a_{kl}(t,x)G^a_{kl}(t,x)\,,
  \label{eq:E_mag_cont}
\end{equation}
are renormalized, i.e.~their expectation values are finite once the standard
renormalizations of the action parameters (gauge coupling and fermion masses)
have been carried out~\cite{Luscher:2011bx}. The gradient flow equation is a non-linear equation in the
$B$-field, however, in perturbation theory to leading order it simply
becomes the heat equation. In 4-momentum space, the $B$- and $A$-fields are
then related by
\begin{equation}
   \tilde{B}_\mu(t,p) = H_{\mu\nu}(t,p,\alpha) \tilde{A}_\mu(p)\,,
\end{equation}
where
\begin{equation}
  H(t,p,\alpha) = \exp(-tK(p,\alpha)),\qquad K_{\mu\nu}(p,\alpha) = p^2 \delta_{\mu\nu} + (\alpha-1)p_\mu p_\nu,
  \label{eq:kernel}
\end{equation}
denotes the free heat kernel, $K$ is the Yang-Mills action kernel 
and $\alpha > 0$ is a parameter to dampen the gauge modes during the flow-time evolution. 
In infinite space-time volume the gauge field propagator at finite flow times $s,t$ takes the form,
\begin{equation}
   \langle B^a_\mu(s,x) B^b_\nu(t,y)\rangle =  \delta^{ab} \int \frac{d^4p}{(2\pi)^4} {\rm e}^{ip(x-y)} \bar{D}_{\mu\nu}(p;s,t;\alpha,\lambda) + {\rm O}(g^2_0) \,,
\end{equation}
where we have passed to the usual perturbative field normalization by rescaling $B\rightarrow g_0B$.
Using a matrix notation with matrix transpose $H^T$ for the Lorentz indices $\mu,\nu=0,\ldots,3$, we have
\begin{equation}
 \bar{D}(p;s,t;\alpha,\lambda) = H(s,p,\alpha) D(p,\lambda) H(t,-p,\alpha)^T,\qquad  D(p,\lambda) = K^{-1}(p,\lambda), 
 \label{eq:Dbar_4mom}
\end{equation}
with gauge parameter $\lambda$. For fixed 4-momentum this propagator is thus obtained by exponentiating
and inverting $4\times 4$ matrices.

We now pass to a finite space-time volume with spatial volume $L^3$ and time extent $T$. 
Assuming $L$-periodic boundary conditions in the spatial directions but not in Euclidean time, the expectation value of the
magnetic energy density is given in terms of the gauge field propagator in a time-momentum representation by
\begin{equation}
  \langle E_{\rm mag}(t,x) \rangle = \frac{N^2-1}{2L^3} \sum_{\bf p} 
  \sum_{k,l=1}^3 \left({\bf p}^2\delta_{kl} - p_k p_l\right)
  \left.\bar{D}_{kl}(x_0,y_0,{\bf p};s,t;\alpha,\lambda)\right\vert_{y_0=x_0;s=t}\,.
  \label{eq:Emag_tree}
\end{equation}
To completely define the propagator for spatial indices we impose either SF boundary conditions,
\begin{equation}
   A_k(x)\vert_{x_0=0} = 0 = A_k(x)\vert_{x_0=T}, \qquad k=1,2,3,
\end{equation}
or SF-open boundary conditions,
\begin{equation}
    A_k(x)\vert_{x_0=0} = 0 = \partial_0 A_k(x)\vert_{x_0=T}, \qquad k=1,2,3,
\end{equation}
and analogously for the $B$-field at positive flow times.

\section{Boundary conditions from an orbifold reflection}

Given a function $\varphi(x_0)$ on a circle of circumference $2T$ one may
obtain boundary conditions at $x_0=0,T$ by introducing a reflection
\begin{equation}
 R:\quad   \varphi(x_0) \rightarrow \varphi(-x_0),
\end{equation}
about $x_0=0$. Since $R$ squares to the identity one may define a corresponding parity and
decompose any such function into even and odd parts,
\begin{equation}
   \varphi(x_0) = \varphi_+(x_0) + \varphi_-(x_0),\qquad   (R\varphi_\pm)(x_0) = \pm \varphi_\pm(x_0).
\end{equation}
It is then easy to see that $\varphi_-(0)=0=\varphi'_+(0)$, i.e.~one obtains Dirichlet and Neumann conditions
at $x_0=0$ for the odd and even parts of $\varphi$, respectively. Moreover, $2T$-periodicity implies
the same boundary conditions at $x_0=T$. Choosing $2T$-antiperiodic boundary conditions 
interchanges the Neumann and Dirichlet conditions at $x_0=T$.

The advantage of inducing boundary condition by such an orbifold reflection lies in the 
remnants of translation invariance of the initial $2T$-(anti)periodic set-up.
In fact, in momentum space,
\begin{equation}
   \varphi(x_0) = \frac{1}{2T}\sum_{p_0} {\rm e}^{ip_0x_0} \tilde{\varphi}(p_0) = 
   \underbrace{\frac{1}{2T}\sum_{p_0}\cos(p_0x_0) \tilde{\varphi}(p_0)}_{\varphi_+(x_0)}
   +\underbrace{\frac{1}{2T}\sum_{p_0}i\sin(p_0x_0) \tilde\varphi(p_0)}_{\varphi_-(x_0)}, 
\end{equation}
the projection onto even and odd components produces the functions $\cos(p_0x_0)$ and $\sin(p_0x_0)$, respectively, 
while leaving $\tilde{\varphi}(p_0)$ unchanged. Boundary conditions at $x_0=T$ depend on the set of momenta being summed over, and the only
difference on the lattice is the finiteness of the sum, and the definition of Neumann conditions by
a lattice derivative.

Applying this reflection principle to the spatial Lorentz components of the gauge field, $\varphi(x_0) = \tilde{B}_k(t,x_0,{\bf p})$ for
$k=1,2,3$, one obtains the propagator in time-momentum representation as a sum over $p_0$,
\begin{equation}
   \bar{D}_{kl}(x_0,y_0,{\bf p};s,t;\alpha,\lambda) = \frac{1}{T} \sum_{p_0} \sin(p_0x_0)\sin(p_0y_0) \bar{D}_{kl}(p;s,t;\alpha,\lambda),
\end{equation}
with the symbol in 4-momentum space as given in eq.~(\ref{eq:Dbar_4mom}). Here the sum is over momenta $p_0$ which 
are allowed by $2T$-periodicity except for $p_0=0$~\cite{Luscher:2014kea},
\begin{equation}
 p_0 = n_0\pi/T,\qquad n_0=\pm 1,\pm 2,\ldots
\end{equation}
or by $2T$-antiperiodicity,
\begin{equation}
   p_0 = (n_0+\tfrac12)\pi/T,\qquad n_0=0,\pm 1,\pm 2,\ldots
\end{equation}
Moreover, taking into account that $\sin(p_0x_0) \sin(p_0 y_0)$ is an even function of $p_0$ one obtains
twice the sum over non-negative values of $n_0$. The Euclidean time-components can be treated similarly but
are not given here, as these are not required for our observable.

\section{Lattice specific expressions}

We refer the reader to ref.~\cite{Ramos:2015baa} for the gradient flow on the lattice
and its Symanzik improvement. Here we are only interested in the perturbative expansion to lowest order, where the
lattice result takes the same form as in eq.~(\ref{eq:Emag_tree}), except that the sums over ${\bf p}$ and $p_0$
have $(L/a)^3$ and $T/a$ terms, respectively, due to the restriction of the momenta to the Brillouin zone,
$-\pi/a \le p_\mu < \pi/a$, for $\mu=0,1,2,3$. Furthermore the Yang-Mills kernels $K(p,\alpha)$ and $K(p,\lambda)$ 
for the flow and the action, respectively, have to be replaced by their lattice counterparts~\cite{Ramos:2015baa}. 
In particular, we need the kernel for the Wilson action,
\begin{equation}
   K^{\rm W}_{\mu\nu}(p,\lambda) = \hat{p}^2\delta_{\mu\nu} +(\lambda -1)\hat{p}_\mu\hat{p}_\nu,
\end{equation}
with the usual lattice momenta $\hat{p}_\mu= (2/a)\sin(ap_\mu/2)$. Slightly more complicated
are the kernels for the L\"uscher-Weisz action,
\begin{equation}
   K^{\rm LW}_{\mu\nu}(p,\lambda) = \hat p^2\delta_{\mu\nu} +
(\lambda-1)\hat p_\mu\hat p_\nu 
   + \tfrac{a^2}{12}\left[ (\hat p^4 + \hat p^2\hat p_\mu^2)\delta_{\mu\nu} - 
   \hat p_\mu \hat p_\nu (\hat p_\mu^2 + \hat p_\nu^2)\right],
\end{equation}
and for the Zeuthen flow, which is obtained from the LW-kernel through,
\begin{equation}
   K^{\rm Z}_{\mu\nu}(p,\alpha) =   \left(1-{\tfrac{a^2}{12}}\hat p_\mu^2\right) K^{\rm LW}_{\mu\nu}(p,0) + \alpha\hat p_\mu\hat p_\nu\,.
\end{equation}

\begin{figure}
\centering
\includegraphics[width=11cm,clip]{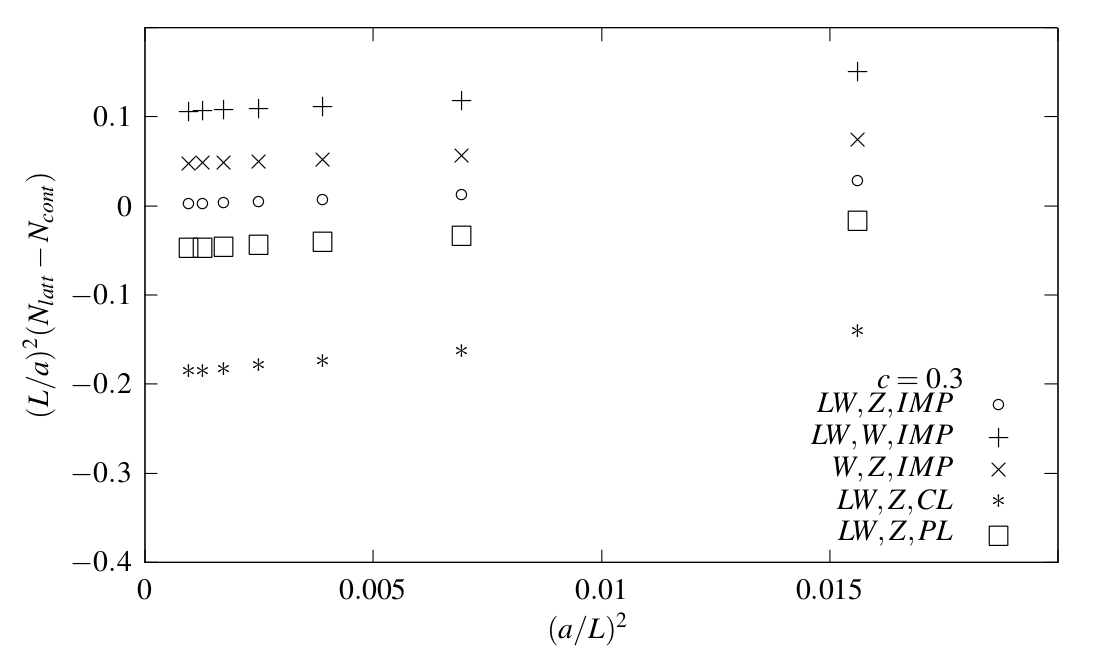}
\caption{The figure shows O($a^2$) cutoff effects with SF boundary conditions at $c=0.3$. 
The labels are ordered in such a way they indicate action, flow, and observable.
$O(a^2)$ improvement is seen only with LW action, Zeuthen flow and improved observable. We then "unimprove" one by one 
to distinguish the cutoff effects origination from the 
Wilson action, Wilson flow and plaquette/clover observables, respectively.}
\label{fig:SF}
\end{figure}

\begin{figure}
\begin{center}
\includegraphics[width=11cm,clip]{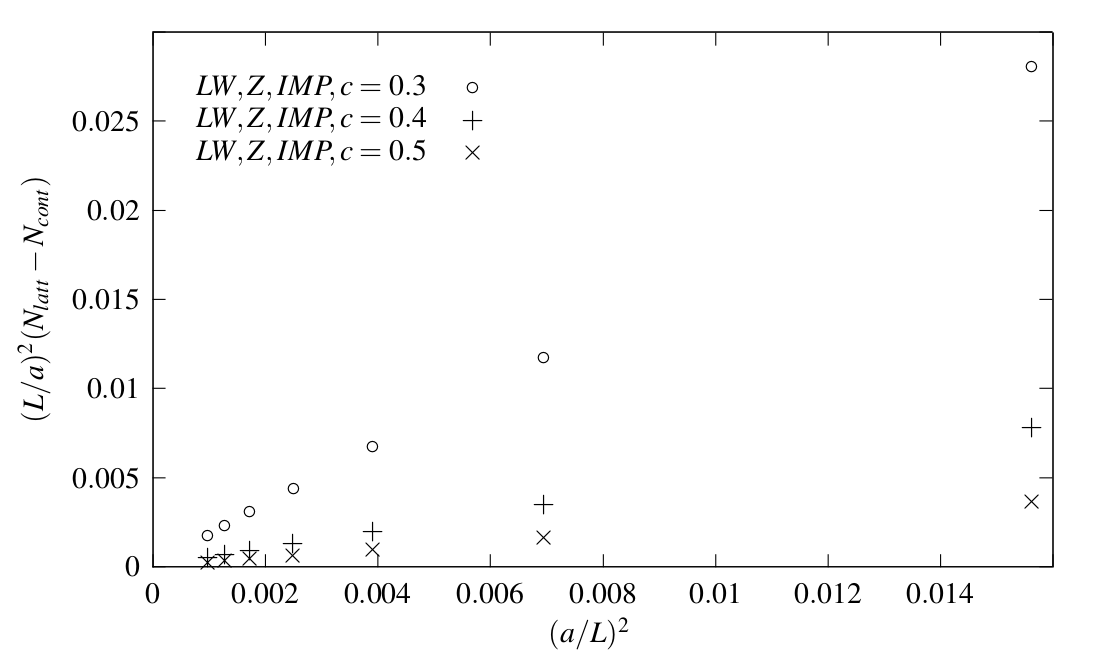}
\caption{O($a^2$) improved data for $c=0.3,0.4,0.5$ and with SF boundary conditions, cf.~text.}
\label{fig:cvalues}
\end{center}
\end{figure}

Finally the lattice observables are obtained using lattice derivatives and thus lattice momenta.
Considering the magnetic energy density, made dimenensionless by a factor $t^2$, as function
of $c= \sqrt{8t}/L$, for $x_0=T/2$, $T=L$  and the lattice resolution $a/L$, we define
\begin{equation}
\label{nnhat}
 t^2 \langle E_{\rm mag}(t,x)\rangle\vert_{c=\sqrt{8t}/L;T=L;x_0=L/2}  = g_0^2\hat{\mathcal{N}} \left(c, \frac{a}{L}\right) + {\rm O}(g_0^4)\,,
\end{equation}
and obtain (with $N=3$ colours),
\begin{equation}
   \hat{\mathcal{N}} \left(c, \frac{a}{L}\right)
 = \frac{c^4}{ 8 } \sum_{ p } \sin^2 \left(\frac{p_0 L}{2}\right)  \sum_{k,l=1}^3 K_{kl}^{\text{3d}}({\bf p}) \bar{D}_{kl}(p)\,,
\end{equation}
where $K_{kl}^{\text{3d}}({\bf p})$ coincides with the reduction to 3 dimensions of the lattice variants of
the kernel $K_{\mu\nu}(p,0)$ (\ref{eq:kernel}), as given in~\cite{Ramos:2015baa}. For instance, the plaquette observable
corresponds to
\begin{equation}
   K^{\text{3d, pl}}_{kl}({\bf p}) = {\bf \hat{p}}^2\delta_{kl} - \hat{p}_k \hat{p}_l\,,
\end{equation}
and the kernel for an O($a^2$) improved observable is then obtained 
as a linear combination with the clover definition~\cite{Ramos:2015baa},
\begin{equation}
 K_{kl}^{\text{3d, imp}}({\bf p}) = \frac43 K_{kl}^{\text{3d, pl}}({\bf p})  - \frac13 K_{kl}^{\text{3d, cl}}({\bf p})\,.
\end{equation}
The numerical evaluation was done with a C++ program  and the Armadillo-library in ref.~\cite{arma} 
for numerical matrix exponentiations and inversions.
We have produced data for the $c$-values $0.3,0.4,0.5$ and lattice resolutions $L/a=8,10,\ldots,32$. 
As we are interested in improvement we look at the difference of to the expected continuum limit (which was also computed numerically).
Rescaling this difference by $(L/a)^2$ and setting $c=0.3$ we observe, in Fig.~\ref{fig:SF}, that indeed only the combination of Zeuthen flow, LW-action and 
improved observable leads to the absence of O($a^2$) effects. Results with SF-open boundary conditions and these parameters look indistinguishable by eye and
are thus not shown. Finally, looking at the O($a^2$) improved setup, we plot the data for SF boundary conditions and 3 $c$-values in Fig.~\ref{fig:cvalues}.
The relevant resolution for cutoff effects should be $a/\sqrt{8t} = a/(cL)$~\cite{DallaBrida:2016kgh}, 
so that one expects a reduction of cutoff effects as $c$ is increased. On the other hand, 
for our choice of $x_0=T/2$ and $T=L$ we expect cutoff effects from the time  boundaries to be no longer exponentially
suppressed once $c$ approaches $0.5$. However, we observe that there are, to this order of perturbation theory, no
visible boundary effects\footnote{O($a$) improvement at the time boundaries is correctly implemented at
tree-level in our set-up.} at O($a^2$) or O($a^3$). This observation also applies to SF-open boundary conditions.
Again, we only show the SF results in Fig.~\ref{fig:cvalues}, as the corresponding plot for SF-open boundary conditions
looks almost identical. While this is expected for $c=0.3$ it remains the case up to $c=0.5$ and despite 
the fact that the continuum limits do show a dependence on the boundary conditions: the discrepancy between SF and SF-open continuum
results for $c=0.3$ is at the 1 per cent level, increasing to about 10 per cent at $c=0.5$.

\section{Conclusion}
We confirm expectations based on the O($a^2$) improved framework of ref.~\cite{Ramos:2015baa}
and previous results in~\cite{DallaBrida:2016kgh}. We have derived 
a convenient representation of the gauge field propagator
using the set-up proposed in \cite{Luscher:2014kea},
which allows us to apply an orbifold reflection principle.  
We anticipate that 
the gauge propagator representation will be very useful in future perturbative
computations which might be needed to complement a non-diagrammatic numerical 
approach~\cite{Brida:2013mva,DallaBrida:2016dai}.

\section*{Acknowledgments}
We thank Alberto Ramos for helpful discussions and for providing an 
independent Fortran code which we used to cross check 
our results with SF boundary conditions.
Financial support by SFI under grant 11/RFP/PHY3218 is gratefully acknowledged.

\end{document}